# Single-objective selective-volume illumination microscopy enables high-contrast light-field imaging

SARA MADAAN,[1,*,†] KEVIN KEOMANEE-DIZON,[1,*] MATT JONES,[2] CHENYANG ZHONG,[1] ANNA NADTOCHIY,[1] PETER LUU,[2] SCOTT E. FRASER,[1] THAI V. TRUONG[2,§]

[1]*Translational Imaging Center, Dornsife College of Letters, Arts and Sciences, and Viterbi School of Engineering, University of Southern California, Los Angeles, CA 90089, USA*
[2]*Translational Imaging Center, Molecular and Computational Biology Section, Department of Biological Sciences, University of Southern California, Los Angeles, CA 90089, USA*
*These authors contributed equally to this work*
[†]*Present address: Google Inc, Mountain View, CA 94043, USA*
[§]*Corresponding author: tvtruong@usc.edu*

**The performance of light-field microscopy is improved by selectively illuminating the relevant subvolume of the specimen with a second objective lens [1-3]. Here we advance this approach to a single-objective geometry, using an oblique one-photon illumination path or two-photon illumination to accomplish selective-volume excitation. The elimination of the second orthogonally oriented objective to selectively excite the volume of interest simplifies specimen mounting; yet, this single objective approach still reduces out-of-volume background, resulting in improvements in image contrast, effective resolution, and volume reconstruction quality. We validate our new approach through imaging live developing zebrafish, demonstrating the technology's ability to capture imaging data from large volumes synchronously with high contrast, while remaining compatible with standard microscope sample mounting.**

Biological processes often depend on the tight spatiotemporal coordination between cells across tissue-level length scales, extending over hundreds of microns in three-dimensions (3D). Functional understanding of such processes would be greatly aided by imaging tools that offer the combined speed and sensitivity needed to observe 3D cellular dynamics without compromising the normal biology [4,5]. Light-field microscopy (LFM) is a fast, synchronous 3D imaging technique [6-8]. Unlike popular volumetric imaging methods that reconstruct a 3D image from intensity information collected one voxel, one line, or one plane at a time, LFM captures both the 2D spatial and 2D angular information of light emitted from the sample [Fig. 1(A)], permitting computational reconstruction of the signal from a full volume in just one shot. Because lateral spatial resolution must be compromised to capture the angular distribution of the emitting light to yield the extended depth coverage, LFM sacrifices some resolution for its dramatically increased acquisition speed. While 3D deconvolution can be used to enhance LFM performance [7,8], out-of-volume fluorescence background, coming from parts of the sample outside of the volume of interest, limits signal detection, image contrast and resolution. Conventional wide-field illumination excites significant out-of-volume background [Fig. 1(B)], especially for volumes within thick or densely fluorescent samples, precluding LFM's full potential in intact tissues.

We recently introduced an improved light-field-based imaging approach, selective-volume illumination microscopy (SVIM), where confining excitation preferentially to the volume of interest reduces extraneous out-of-volume background, thereby sharpening image contrast, reducing unwanted photodamage, and improving the *effective* resolution in thick specimens (hundreds of microns or more) [1,3]. SVIM was implemented with two objective lenses: one to selectively illuminate the volume of interest, and a second objective, orthogonally aligned, to acquire the fluorescent light-field [Fig. 1(C)]. This two-objective geometry requires that the sample be mounted within the mutual intersecting volume defined by the perpendicular objectives, complicating sample mounting and limiting sample size. Here, we implement SVIM in a single-objective geometry, eliminating the need for two orthogonally oriented objectives, greatly simplifying sample mounting and broadening its utility for biological research.

This new technique, termed axial single-objective SVIM (ASO-SVIM), selectively illuminates the sample volume through the same objective used for high-numerical-aperture (NA) detection (Supplement 1, Section 1, and Fig. S1). The volume of interest is preferentially excited by either one-photon or two-photon processes (1P- or 2P-ASO-SVIM). 1P-ASO-SVIM is accomplished by using a 2D light-sheet oriented obliquely to the axial axis [Fig. 1(D)], created via a cylindrical lens; the sample is illuminated by sweeping this oblique sheet in 1D to excite fluorescence within the desired region of interest, multiple times within a single camera exposure. 2P-ASO-SVIM [Fig. 1(E)] is accomplished using a low-NA Gaussian beam that is raster-scanned in 2D to excite the 3D sample volume of interest.

To capture fluorescence light-fields emitted from the excited volume, a lenslet array is placed at the native image plane [7]; the foci of the lenslets are imaged onto a camera sensor [Fig. 1(A)]. To enable direct, quantitative comparison of our technique to more established methods, our microscope is designed to offer seamless switching to light-sheet (also known as selective-plane illumination microscopy; SPIM) or wide-field LFM modes [Supplement 1, Section 1, and Figs. S1-S2].

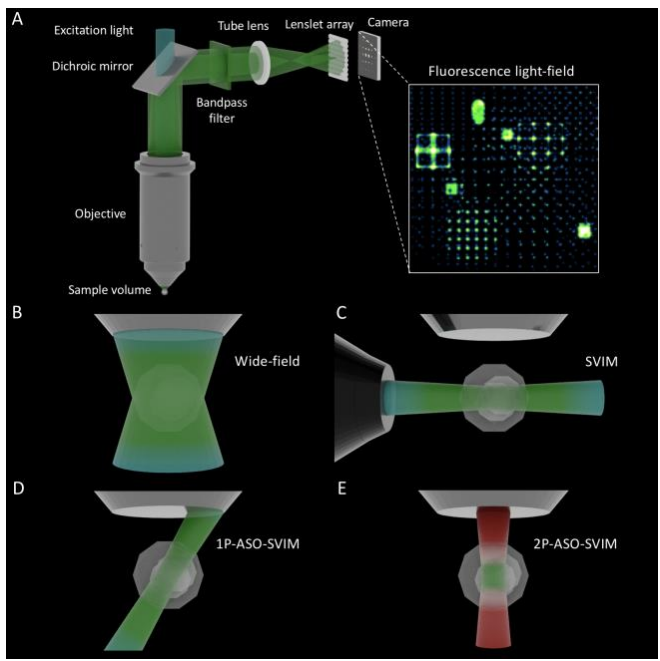

Fig. 1. Axial single-objective selective-volume illumination (ASO-SVIM). (A) Simplified schematic of light-field microscopy (LFM). Fluorescence light is collected from the sample volume by an objective lens, separated and filtered from the excitation light by the appropriate dichroic mirror and bandpass filter, and focused by a tube lens at an intermediate image plane where a lenslet array is positioned. The lenslet array refocuses the light onto a camera, so that each position in the 3D sample volume is mapped onto the camera as a unique light intensity pattern. The fluorescence light-field illustrated was captured with point-sources located at, above, and below the native focal plane. Such light-fields can be reconstructed to full volumes by solving the inverse problem [3].
(B) LFM with conventional wide-field illumination is compatible with standard forms of sample preparation but excites regions outside of the volume of interest (VOI).
(C) Inspired by light-sheet microscopy (SPIM), SVIM selectively illuminates the VOI using orthogonal illumination and detection objectives. Shown previously [1-3], SVIM reduces background fluorescence outside the VOI, increasing image resolution and contrast.
(D, E) ASO-SVIM preferentially excites the VOI and collects the fluorescence using a single objective lens, providing flexibility in sample mounting similar to traditional microscopy. (D) 1P-ASO-SVIM uses an oblique light sheet, that is scanned in 1D, to define the excitation volume. (E) 2P-ASO-SVIM uses nonlinear excitation of a pulsed near-infrared (NIR) beam that is raster-scanned to define the VOI.
In each figure, 1P excitation is depicted in cyan (A-D) and 2P excitation in red (E); fluorescence emission is depicted in green.
See also Supplement 1, Section 1, and Figs. S1-S2.

We benchmarked ASO-SVIM performance by measuring the point-spread function (PSF) with 175-nm fluorescent beads suspended in agarose. After 3D deconvolution [7,8], we obtained volumetric images with the expected maximum resolution, consistent with the optical design: $2.4 \pm 0.3$ μm lateral full-width at half-maximum (FWHM); $5.7 \pm 0.2$ μm axial FWHM [Supplement 1, Fig. S3(C)]. Due to diffraction and non-uniform sampling of the light-field volume [7,8], the 3D resolution was depth-dependent (varying up to ~46 % over a $z$ range of -50 to 50 μm) [Supplement 1, Fig. S3(B)], and reconstructions contained grid-like artifacts near the native focal plane, as previously reported [7]. To reduce such artifacts in the reconstructions presented here, we applied a low-pass filter in Fourier space ($k$-space), truncating spurious spatial frequencies beyond the resolution limit of the native focal plane (Supplement 1, Section 2, Figs. S4-S5). The simple process of $k$-space filtering across the nominal focus dampened the artifacts and improved visualization of the 3D reconstructions, without any major loss of 3D resolution or spatial information (Supplement 1, Figs. S5-S6).

To test ASO-SVIM on biological samples, we imaged the vasculature of live larval zebrafish (at 5 days post-fertilization, dpf), labeled with green fluorescent protein (GFP). Zebrafish embryos and larvae are ideal for studies involving multicellular and multiscale imaging because of their small size, transparency, and amenability to genetic engineering. As expected, ASO-SVIM, using either 1P or 2P excitation, produced less out-of-volume background than did wide-field illumination [Fig. 2(A)], and this reduced background fluorescence yields higher contrast images, as we previously reported for SVIM [1,3]. This is clearly revealed in an $x$-$z$ slice through the 3D volume [Fig. 2(A), bottom]. Although ASO-SVIM reconstructed images all fell short of the quality of the ground truth images (a deconvolved SPIM image stack), all three dimensions were acquired simultaneously, generating the 3D image more than 100-fold faster.

To obtain quantitative measures of the enhanced performance of 1P-ASO-SVIM and 2P-ASO-SVIM, we calculated the image contrast (defined as the standard deviation of the pixel intensities normalized to the mean intensity) for each $x$-$y$ image plane [Fig. 2(B)]: 1P-ASO-SVIM showed a modest improvement; 2P-ASO-SVIM showed more dramatic improvement over wide-field LFM. Intensity profiles of LFM images [along the dashed yellow line in Fig. 2(A), bottom] documented the improved performance of ASO-SVIM [Fig. 2(C)], as did Fourier transforms of the images [Fig, 2(D,E)]. Thus, the reduction in background fluorescence substantially boosts image contrast as well as effective spatial resolution, both laterally and axially [Fig. 2(D,E) and Supplement 1, Fig. S7]. The enhanced contrast and effective resolution of the ASO-SVIM modalities are further demonstrated by comparing recorded images of single blood vessels, as well as measurements of the width of these vessels from the line profiles (Fig. 2(F,G) and Supplement 1, Fig. S8).

To test ASO-SVIM on a more demanding application, we recorded the activity of large populations of neurons in larval zebrafish. Imaging the nervous system in action within the intact brain is challenging because it requires cellular resolution over thousands of cells with sufficient temporal resolution to capture the transient firing of neurons. LFM is an attractive technique to meet these neuroimaging challenges because it can synchronously capture large volumes; however, the high level of background fluorescence in wide-field LFM has remained an impediment to efforts aimed at capturing brain-wide activity with cellular resolution [8,10-16]. We previously showed that the improved contrast and effective resolution of SVIM improved brain-wide functional imaging over conventional LFM [3]. We extended the demonstration and analysis to our new ASO-SVIM approach here.

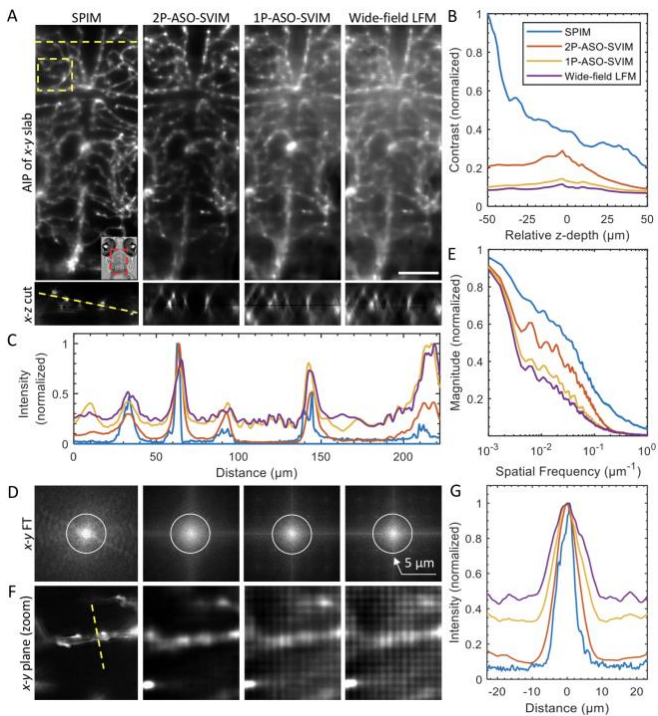

Fig. 2. ASO-SVIM improves contrast and effective resolution in live imaging of zebrafish larvae.

(A, Top) Average-intensity projections (AIPs) of a 100-μm-thick 3D image stack from the same transgenic [Tg(kdrl:GFP)] 5-dpf zebrafish, where the vasculature was fluorescently labeled, captured by different imaging modalities.

(Inset) Transmitted light image of the zebrafish head, where the dashed red rectangle marks the 230 μm × 600 μm × 100 μm volume imaged.

(Bottom) Cross-sectional (x-z) views at the location indicated by the dashed yellow line (top left, SPIM panel).

The SPIM volume consisted of 67 optical sections, collected serially over ~44 s; the LFM-based volumes were reconstructed from a single image with an exposure of 355 ms. Scale bar, 100 μm.

(B) Quantification of image contrast versus z-depth, showing improvements of 1P-ASO-SVIM and 2P-ASO-SVIM over wide-field LFM. Image contrast, measured as the standard deviation of the pixel intensities divided by the mean intensity, is plotted for each x-y slice, normalized by the value of the deconvolved SPIM (blue trace) at z = -50 μm.

(C) Intensity profiles along the same line for all four modalities (dashed yellow line shown on the x-z cut in A, bottom, SPIM). The fluorescence intensities of ASO-SVIM better agree with the ground truth SPIM data than does wide-field LFM.

(D) Fourier transforms (FTs) of x-y MIPs through the 100-μm-thick slabs in (A). Resolution bands (white circles) help show the increased spatial frequency content of ASO-SVIM compared to wide-field LFM, where more signal intensity at larger radial position designates higher resolution captured.

(E) Average amplitudes along $k_y$ direction of the FTs shown in (D), showing that ASO-SVIM frequency spectra fall slower to the experimental noise floor, indicating better effective resolution than wide-field LFM. See also Supplement 1, Fig. S7.

(F) Enlarged x-y slices to highlight single blood vessels, centered at ~86 μm into the specimen, from the subregion indicated by dashed yellow box in (A, top left).

(G) Intensity profiles, averaged from the dashed yellow line shown in (F) and from 2 other similar vessel structures (Supplement 1, Fig. S8), demonstrate the improvements in effective resolution of ASO-SVIM over wide-field LFM, in measuring the width of blood vessels.

To permit direct comparisons between modalities, we used 1P-ASO-SVIM, 2P-ASO-SVIM, and wide-field LFM to image the spontaneous brain activity of the same 5-dpf zebrafish, labeled with a genetically encoded pan-neuronal fluorescent calcium indicator [Fig. 3]. The reconstructed 4D recordings are compared by taking the standard deviation along the temporal axis [Fig. 3(A)], to highlight their capability in capturing active neurons, whose transient firings would produce voxels that have large intensity variation in time and thus appear as high-contrast puncta in the resulting projections. We calculated the image contrast of these temporal-standard-deviation projections: 2P-ASO-SVIM achieved the highest contrast, followed by the 1P ASO-SVIM, and then by wide-field LFM [Fig. 3(B)], suggesting that the ASO-SVIM modalities excel in capturing neuronal activity over wide-field LFM.

To quantitatively compare the performance of the different modalities in capturing brain activity at cellular resolution, we identified neurons in the 4D recordings by spot-segmenting the temporal-standard-deviation projections. This standard protocol [3] produced spatial masks corresponding to neurons that were active during the time-lapse. These masks were then applied to the 4D datasets to extract temporal signals that represent single-neuron activity traces [Fig. 3(C)]. The improved contrast of 2P-ASO-SVIM and 1P-ASO-SVIM allowed us to detect a greater number of active neurons in the brain compared to conventional wide-field illumination [Fig. 3(C)]. 2P-ASO-SVIM captured the largest number of active neurons, due not only to its higher contrast than its 1P counterpart (expanded below) but also because the NIR excitation light is invisible to the fish and thereby significantly reduces the response of the animal's visual system to the illumination, which would otherwise cloud spontaneous activity [3]. 2P-ASO-SVIM is thus an optimal tool for studies of visually sensitive neural behaviors.

1P-ASO-SVIM and 2P-ASO-SVIM offer distinct strengths. 1P-ASO-SVIM commands lower laser costs, and offers optical simplicity and exceptionally high volumetric acquisition speed, limited largely by the rate of the camera [3]. However, the 1P excitation volume is larger and intersects the sample obliquely [Fig. 1(D)], making 1P-ASO-SVIM less efficient at reducing background than SVIM. Like all forms of linear excitation, visible 1P excitation light increasingly scatters with depth, resulting in unavoidable background from outside the volume of interest. 2P-ASO-SVIM effectively eliminates background from out-of-volume fluorescence [Fig. 2(a)] due to nonlinear excitation: The quadratic dependence of 2P-excited fluorescence on the laser intensity restricts the excitation volume to near the focus [3,9], resulting in negligible background even with single-objective designs. The NIR excitation light is scattered much less than visible wavelengths, and any scattered light is unlikely to generate background as it is unlikely to achieve the intensity required to excite fluorescence or autofluorescence in tissue. Through the judicious selection of illumination NA and beam-scanning, it is straightforward to match the 2P excitation volume to the desired light-field region of interest (Supplement 1, Section 1). This advantage is partially tempered by the reduced speed of 2P-ASO-SVIM, as the lower 2P excitation cross section yields lower fluorescence signal for a given laser intensity, which cannot be increased without bounds out of concern for photodamage.

As a final example of the combination of high-contrast, ultrahigh-speed volumetric imaging at cellular resolution and the sample-mounting flexibility of ASO-SVIM, we imaged 3D blood flow in nearly the entire larval zebrafish brain, covering a 670 μm × 470 μm × 200 μm volume at ~50 Hz, in 9 zebrafish mounted in a standard multi-well plate (Supplement 1, Fig. S9, and Visualizations 2-4). Together, our results show that ASO-SVIM offers a convenient middle ground between SPIM and traditional wide-field LFM, offering improved contrast and effective resolution compared to LFM, while outperforming the 3D imaging speed of SPIM by approximately two orders of magnitude, as it requires only a single camera exposure to capture an extended volume. Compared to our earlier form of SVIM [1, 3], ASO-SVIM relaxes steric constraints by using only one objective, similar to recent developments in single-objective light-sheet-based microscopy [18-21], easing sample preparation and expanding the application space to multicellular systems that are impractical for a dual-objective design. Finally, the simplicity of ASO-SVIM renders it compatible and synergistic with many recent refinements of LFM [10-13], and we envision that together they will bring LFM-based imaging techniques into a wide range of biological systems and applications.

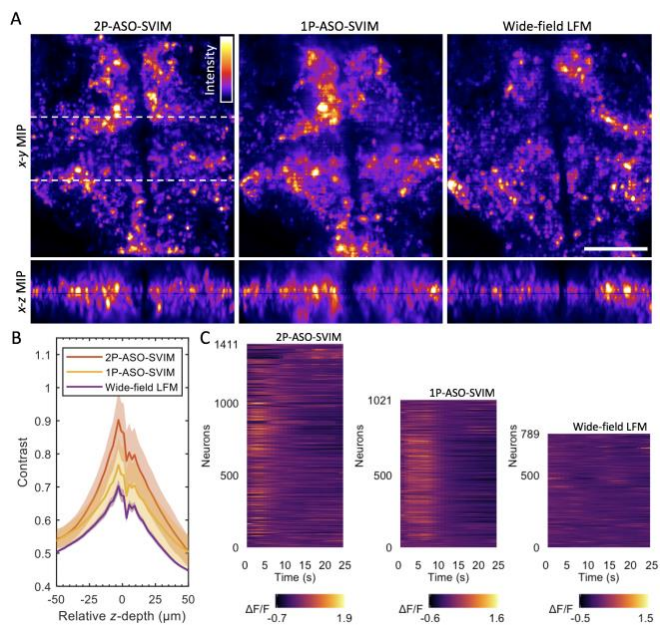

Fig. 3. ASO-SVIM enhances large-scale *in vivo* recording of neural activity in a 320 μm × 350 μm × 100 μm volume in larval zebrafish, at 5-dpf, Tg(elavl3:H2b-GCaMP6s). Volumetric rate of 1 Hz sufficiently captured the transient neuronal dynamics given the relatively slow temporal kinetics of the nulcear-localized calcium indicator GCaMP6s [17].
(A) MIPs of *x-y* (top) and *x-z* (bottom) brain-wide 100-μm-thick volumes of the same zebrafish. These projections depict the standard deviation over a 25-s period for each voxel, highlighting active neurons. Scale bar, 100 μm. See Visualization 1.
(B) Quantification of image contrast versus *z*-depth, showing progressive improvements of 1P-ASO-SVIM and 2P-ASO-SVIM over wide-field LFM. Means (center lines) and standard deviations (shadings) are shown.
(C) Single-neuron signal traces captured by the different modalites, extracted from the 4D datasets shown in (A). The greatest quantity of neurons were found with 2P-ASO-SVIM, followed by 1P-ASO-SVIM, and then wide-field LFM. See text for description of signal extraction protocol.

**Funding.** US National Science Foundation (1608744, 1650406, 1828793); US National Institutes of Health (1R01MH107238-01); Human Frontier Science Program (RGP0008/2017); Jet Propulsion Laboratory Research Subcontract (1632330); Alfred E. Mann Doctoral Fellowship (to KKD).

**Acknowledgment**. We thank Dan Holland, Andrey Andreev, and Francesco Cutrale for insightful discussions.

**Competing interests**. TVT, SM, and SEF: patent application number PCT/US2017/019512 (pending).

See Supplement 1 for supporting content.

## REFERENCES

1. T.V. Truong, D.B. Holland, S. Madaan, A. Andreev, J.V. Troll, D.E.S. Koo, K. Keomanee-Dizon, M.J McFall-Ngai, and S.E. Fraser, bioRxiv (2018).
2. N. Wagner, N. Norlin, J. Gierten, G. de Medeiros, B. Balázs, J. Wittbrodt, L. Hufnagel, and R. Prevedel, "Instantaneous isotropic volumetric imaging of fast biological processes," Nat. Methods **16**, 497 (2019).
3. T.V. Truong, D.B. Holland, S. Madaan, A. Andreev, K. Keomanee-Dizon, J.V. Troll, D.E.S. Koo, M.J McFall-Ngai, and S.E. Fraser, Comm. Biol. **3**, 1 (2020).
4. W. Supatto, T.V. Truong, D. Débarre, E. Beaurepaire, Cur. Op. Gen. Dev. **21**, 538 (2011).
5. F. Cutrale, S.E. Fraser, and L.A. Trinh, Annual Rev. of Biomed. Data Science **2**, 223 (2019).
6. M. Levoy, R. Ng, A. Adams, M. Footer, and M. Horowitz, in *Proceedings of ACM SIGGRAPH* (ACM, 2006) p. 924.
7. M. Broxton, L. Grosenick, S. Yang, N. Cohen, A. Andalman, K. Deisseroth, and M. Levoy, Opt. Express **21**, 25418 (2013).
8. R. Prevedel, Y. G. Yoon, M. Hoffmann, N. Pak, G. Wetzstein, S. Kato, T. Schrödel, R. Raskar, M. Zimmer, E. S. Boyden, and A. Vaziri, Nat. Methods **11**, 727 (2014).
9. T.V. Truong, W. Supatto, D.S. Koos, J.M. Choi, and S.E. Fraser, Nat. Methods **8**, 757 (2011).
10. N.C. Pégard, H.Y. Liu, N. Antipa, M. Gerlock, H. Adesnik, and L. Waller, Optica **3**, 517 (2016).
11. L.M. Grosenick, M. Broxton, C.K. Kim, C. Liston, B. Poole, S. Yang, A.S. Andalman, E. Scharff, N. Cohen, O. Yizhar, C. Ramakrishnan, S. Ganguli, P. Suppes, M. Levoy, and K. Deisseroth, bioRxiv, 132688 (2017).
12. L. Cong, Z. Wang, Y. Chai, W. Hang, C. Shang, W. Yang, L. Bai, J. Du, K. Wang, and Q. Wen, eLife **6**, e28158 (2017).
13. T. Nöbauer, O. Skocek, A.J. Pernía-Andrade, L. Weilguny, F.M. Traub, M.I. Molodtsov, and A. Vaziri, Nat. Methods **14**, 811 (2017).
14. O. Skocek, T, Nöbauer, L. Weilguny, F.M. Traub, C.N. Xia, M.I. Molodtsov, A. Grama, M. Yamagata, D. Aharoni, D.D. Cox, P. Golshani, and A. Vaziri, Nat. Methods **25**, 429 (2018).
15. S. Aimon ,T. Katsuki, T. Jia, L. Grosenick, M. Broxton, K. Deisseroth, T.J. Sejnowski, and R.J. Greenspan, PLoS Biol. **17**, e2006732 (2019).
16. Q. Lin, J. Manley, M. Helmreich, F. Schlumm, J.M. Li, D.N. Robson, F. Engert, A. Schier, T. Nöbauer, and A. Vaziri, Cell **180**, 536 (2020).
17. T.-W. Chen, T. J. Wardill, Y. Sun, S. R. Pulver, S. L. Renninger, A. Baohan, E. R. Schreiter, R. A. Kerr, M. B. Orger, V. Jayaraman, L. L. Logger, K. Svoboda, and D. S. Kim, Nature **499**, 295 (2013).
18. V. Voleti, K.B. Patel, W. Li, C.P. Campos, S. Bharadwaj, H. Yu, C. Ford, M.J Casper, R.W. Yan, W. Liang, C. Wen, K.D. Kimura, K.L. Targoff, E.M.C. Hillman, Nat. Methods **16**, 1054 (2019).
19. B. Yang, X. Chen, Y. Wang, S. Feng, V. Pessino, N. Stuurman, N.H. Cho, K.W. Cheng, S.J. Lord, L. Xu, D. Xie, R.D. Mullins, M.D. Leonetti, and B. Huang, Nat. Methods **16**, 501 (2019).
20. J. Kim, M. Wojcik, Y. Wang, S. Moon, E.A. Zin, N. Marnani, Z.L. Newman, J.G. Flannery, K. Xu, and X. Zhang, Nat. Methods **16**, 853 (2019).
21. E. Sapoznik, B.J Chang, R.J. Ju, E.S. Welf, D. Broadbent, A.F. Carisey, S.J. Stehbens, K.M. Lee, A. Marín, A.B. Hanker, J.C. Schmidt, C.L. Arteaga, B. Yang, R. Kruithoff, D.P. Shepherd, A. Millett-Sikking, A.G. York, K.M. Dean, R. Fiolka, bioRxiv 030569 (2020).

# Single-objective selective-volume illumination microscopy enables high-contrast light-field imaging: supplementary material

# 1. Microscope optics

We describe here the light-field-based selective-volume illumination microscope used in our work. Refer to Fig. S1 for the beam paths and key components.

*1.1 ASO-SVIM: oblique-angled one-photon excitation and wide-field illumination modes*

The illumination path for one-photon (1P) excitation, represented by the blue line, is provided by a bank of continuous-wave (CW) fiber lasers (Coherent OBIS LX, UFC Galaxy: 488 nm, 30 mW; 514 nm, 50 mW; 640 nm, 75 mW) and high-power CW lasers (488 nm, 300 mW, Coherent Sapphire LP; and 532 nm, 5 W, Coherent Verdi). Light from the CW laser bank is collimated and expanded by an objective (BE; Nikon, Plan Fluorite 10×, 0.3 NA, 16 mm WD), directed by a dichroic mirror (DC1; FF750-SDi02-25x36, Semrock), and passed through a remote refocus module, which is composed of lens pair $T1_1$ and $T1_2$ (both 75-mm focal length, Thorlabs AC254-075-A-ML). Adjusting the position of $T1_2$ refocuses the beam waist so that it is coincident to the nominal detection focal plane at the sample. The illumination beam is then sent to a 2D (*x*-*y*) scanning galvo system (G; 6-mm aperture silver mirrors, Cambridge Technology H8363) before being passed through a scan lens (SL; 110-mm focal length, Thorlabs LSM05-BB), a tube lens (TL; 150-mm focal length, Thorlabs AC508-150-B), and a water-dipping objective (ASO; Nikon, CFI LWD Plan Fluorite 16×, 0.8 NA, 3 mm WD); G, SL, and TL are mounted on a computer-controlled motorized translational stage (Newport 436 and Newport LTA-HS) to control the inclination angle in ASO-SVIM mode (tilted 26.5° relative to the optical axis of ASO; purple dashed line), and easily port the beam back to the wide-field illumination mode. The illumination NA is adjusted to be ~ 0.04 to 0.06, depending on the selective illumination extent, yielding a fluorescence Gaussian-beam waist of ~ 4 to 6 μm with an axial (*z*) extent ranging from ~150 to 230 μm (measured as the confocal parameter of the focal volume). As G is conjugate to the back pupil of ASO, scanning along the *x*- and *y*-axes with the appropriate voltages selectively paints out the desired sample volume.

For fast volumetric 1P imaging, the high-power CW laser was used to provide the high laser intensity needed beyond what the CW laser bank could provide. Light from the high-power CW lasers are collimated and expanded by BE (Thorlabs BE052-A) and directed by mirrors to a cylindrical beam-shaping module, composed of a pair of cylindrical lenses $C_1$ and $C_2$ (-50-mm focal length, Thorlabs LK1662L1 or -30-mm focal length, Thorlabs LK1982L1; and 150-mm focal length, Thorlabs LJ1629L1) which expand the beam elliptically in the *y*-direction. This expanded beam is reflected by a mirror mounted on a motorized motion-control stage (MM1; Newport 436 and Newport LTA-HS), where it is directed through $T1_1$ and $T1_2$ and then focused into a 2D (*y*-*z*) sheet by $C_3$ (75-mm focal length, Thorlabs LJ703RM-A) onto G. Thus, G only needs to provide scanning along the *x*-axis to selectively paint out the desired volume at the sample. Note that $C_3$ is used only for 1D scanning, and omitted in the other imaging modes. All 1P imaging data were acquired with 1D scanning except for Fig. 3, where 2D scanning was employed to provide a more precise selectively-illuminated volume, in order to avoid direct illumination of the animal's eyes. An inspection camera (not shown; PCO pco.edge 5.5) conjugate to the sample volume and coincident to the *x*-*z* plane aid in alignment and calibration of the illumination tilt angle and G scanning parameters. Tradeoffs associated with volume-scanning as well as alternative implementations of selective-volume illumination are discussed in [1].

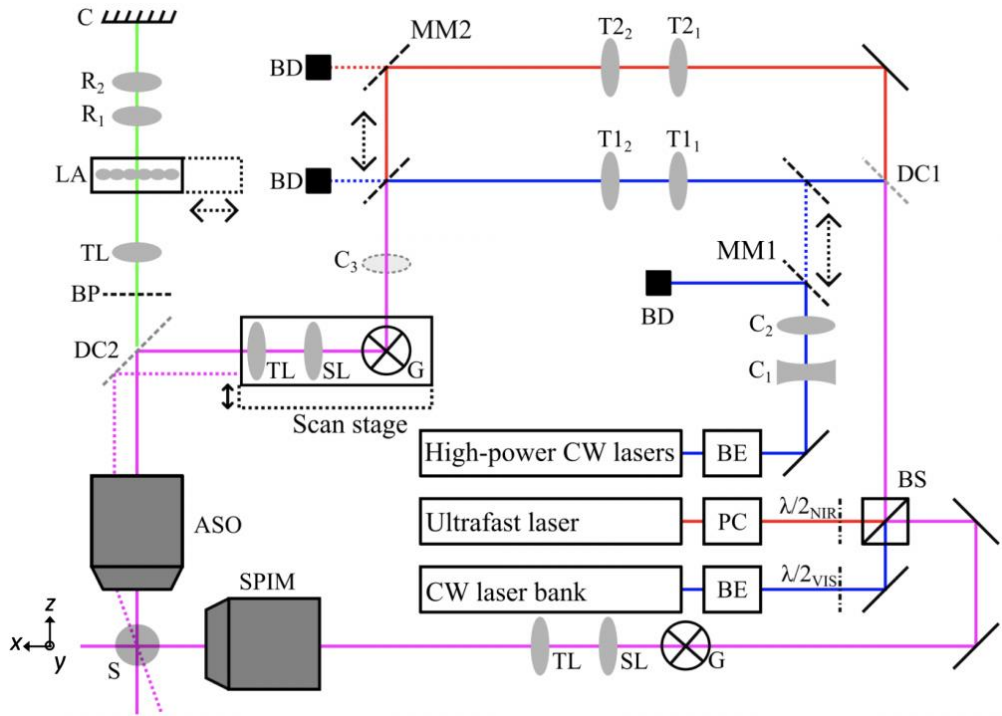

Fig. S1. Simplified schematic diagram of ASO-SVIM. CW: continuous-wave; BE: beam expander; PC: Pockels cell; λ/2: half-wave plate, where the subscripts VIS and NIR refer to the visible and near-infrared wavelengths, respectively; BS: polarizing beamsplitter; DC: dichroic mirror, T1: VIS relay lens, T2: NIR relay lens, C: VIS cylindrical lens, MM: movable mirror, where the subscripts refer to the sequence of elements; BD: beam dump; G: 2D scanning galvo mirrors; SL: scan lens; TL: tube lens; ASO: ASO-SVIM objective; SPIM: light-sheet excitation objective; S: sample; BP: band-pass filter; LA: lenslet array; R: detection relay lens, where the subscripts refer to the sequence of lenses; C: camera sensor.

*1.2 2P-ASO-SVIM: two-photon excitation mode*

The illumination path for two-photon (2P) excitation begins in red. Near-infrared (NIR) pulsed illumination is provided by a Ti:Sapphire ultrafast laser (Coherent Chameleon Ultra II) and the illumination power is controlled by a Pockels cell (PC; Conoptics 350-80). A polarizing beamsplitter (PBS; Thorlabs PBS102) is used to combine the visible and NIR beams into a co-linear beam and to split the combined beam into two integrated excitation paths (towards ASO and SPIM objectives). Visible and NIR half-wave plates ($\lambda/2_{VIS}$ and $\lambda/2_{NIR}$; Thorlabs AHWP05M-600 and AHWP05M-980), each mounted in manual rotation mounts, are used to adjust the laser power delivered to ASO and SPIM as appropriate. In the ASO path, the NIR illumination beam is transmitted through DC1 and then through lens pair $T2_1$ and $T2_2$ (75-mm focal length, Thorlabs AC254-100-B and 100-mm focal length, Thorlabs AC254-75-B-ML), used to expand and refocus the beam waist before being sent to the same illumination-scanning optics in the aforementioned 1P mode (G, SL, TL, and ASO). A mirror mounted on a motion-control stage (MM2) allows automated switching between 2P- and 1P-ASO excitation. The illumination NA is adjusted to be ~ 0.055 to 0.08, yielding similar fluorescence Gaussian-beam characteristics as the 1P mode: ~ 4 to 5 μm waist and ~150 to 230 μm axial extent. For all 2P imaging experiments presented (Figs. 2-3), ~ 525 mW of average laser power was delivered to the specimen.

## 1.3 Light-field detection and reconstruction

Excited fluorescence at the sample is collected by the ASO objective. A dichroic mirror (DC2; Di01-R488/561 or di01-R405/488/543/635-25x36) and a filter wheel (Sutter Instrument Lambda 10-3, 32 mm diameter) equipped with emission filters (FF01-470/28-32, FF03-525/50-32, FF01-609/54-32, and FF01-680/42-32) together block the excitation light and transmit the fluorescence signal emitted from the sample (green). An intermediate image at an overall magnification of 24× is projected onto a lenslet array (LA; 2.06-mm focal length, 18x18 mm, 136 μm pitch, AR coated, OKO Technologies APO-Q-P192-F3.17; f-number matched to the NA of ASO) by a tube lens (TL; 300-mm focal length, Edmund Optics 88-597). With LA placed at the native image plane, an array of fluorescence focal spots is created, which encode 4D spatio-angular information for each position in the 3D volume—referred to as the light-field [3,4]. The generated light-field is imaged onto an sCMOS camera (C; Andor Zyla 5.5) by a pair of photographic lenses $R_1$ and $R_2$ (both 50-mm focal length, Nikon NIKKOR f/1.4). These raw light-fields are reconstructed into full volumes as described in refs. [1,5]. Unless otherwise noted, all image stacks are further processed using a filtering algorithm described in Section 2.

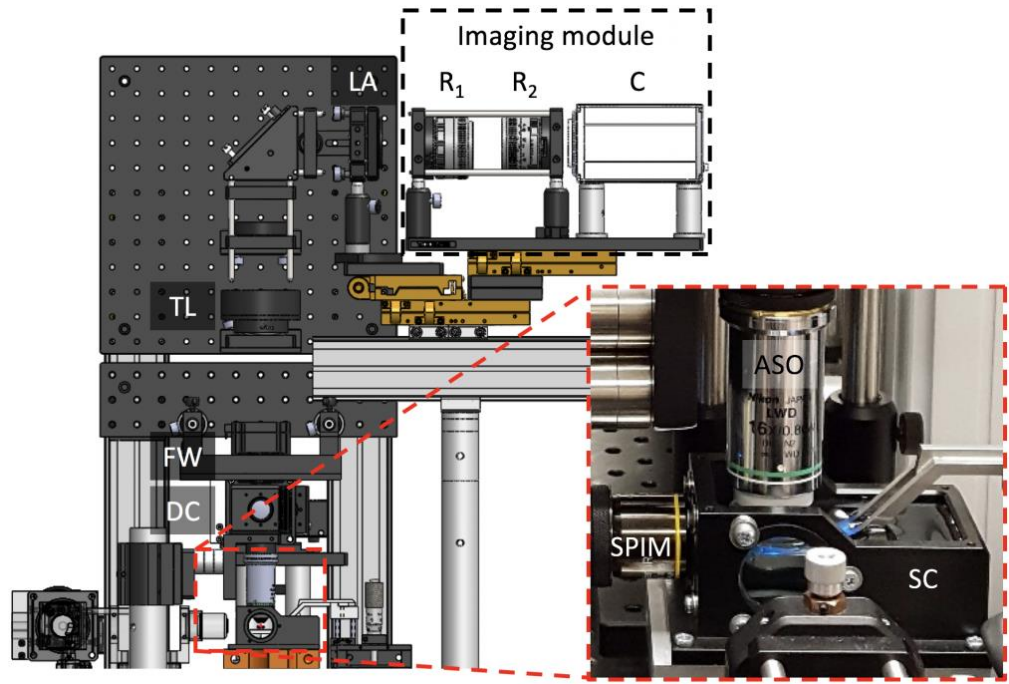

Fig. S2. 3D opto-mechanical model of the ASO-SVIM light-field detection path. Inset shows a photograph of the sample chamber, the axial-single-objective (ASO) used to both deliver selective-volume illumination at the sample and collect the excited fluorescence, as well as the light-sheet excitation objective (SPIM). Owing to the ASO design, samples can be mounted using a caddy and dive bar system as described in ref. [2] and are entirely compatible with standard sample preparation protocols (e.g., Fig. S9). Fluorescence collected from ASO passes through a dichroic mirror (DC), a filter wheel (FW), a tube lens (TL), a lenslet array (LA), and onto an imaging module. R: detection relay lens, where the subscripts refer to the sequence of lenses; C: camera.

## 1.4 SPIM: one-photon and two-photon light-sheet imaging modes

In order to operate in SPIM mode, either $\lambda/2_{VIS}$ or $\lambda/2_{NIR}$ is rotated so that enough excitation energy is transmitted through PBS and delivered at the sample. After PBS, the illumination beam is routed to a 2D (x-z) scanning galvo system (G; 5-mm aperture silver mirrors, Thorlabs GVSM002), and then passed through SL, TL, and an objective (SPIM; Olympus, LMPLN-IR

10×, 0.3 NA, 18 mm WD) to excite the sample with a scanned Gaussian-beam light-sheet. The SPIM objective is mounted on a manual translational stage to create more sample space for ASO-SVIM mode if needed.

In order to collect images in SPIM mode, LA is moved entirely out of the detection path, and the entire imaging module (R1, R2, and C) is moved in -$z$ by the focal length of LA. As shown in Fig. S2, LA and the imaging module are each mounted on motorized linear translational stages (Newport 436 and Newport LTA-HS), enabling high-precision positioning and seamless switching between light-field and conventional wide-field/SPIM detection via computer command. The stages also serve to aid in fine alignment. To assemble a 3D volume, 2D images are recorded in series by scanning the sample in $z$ through the stationary light-sheet with a motorized stage (Newport 436 and Newport LTA-HS).

### 1.5 Instrument control

Instrument control is similar to our previous implementation [1], with the primary changes concerning the coordination between the scanning system and camera triggering. In our new single-objective configuration, a combination of custom software developed in LabView (National Instruments), ScanImage [6], and Micro-Manager [7] synchronize the scanning system, laser intensity, and camera triggering so that the volume of interest is illuminated an integer number of times within one camera exposure and the excitation intensity is near-uniform frame-to-frame during acquisition. All the motorized linear translational stages used to switch between modes are controlled by an XPS Universal Motion Controller (Newport XPS-Q8). The 3D stage stack-up (Sutter MP-285) used for sample positioning is controlled with its corresponding controller; the sample-scanning $z$-stage (noted in Section 1.4) is controlled via Micro-Manager.

### 1.6 Characterizing system resolution

To quantify resolution in volumetric reconstructions of light-fields, we measured the point-spread function (PSF) with 175-nm fluorescent beads sparsely suspended in agarose (Fig. S3). We stepped the sparse bead sample in $z$ by 2 μm over a 200-μm volume, imaging the same field of beads at different axial depths, and thereby facilitating multiple measurements of isolated beads throughout the light-field volume. The $z$-series of light-field images were then reconstructed to yield a series of 3D-stacks with overlapping $z$-extents, from which we calculated the resolution as a function of relative $z$-depth (Fig. S3B). The observed relation between relative $z$-depth and the PSF are consistent with results derived from wave optics theory [4]: at different axial depths, the PSF size is different, generally broadening away from the native focus symmetrically; on the other hand, bead-measured PSFs across reconstructed 2D ($x$-$y$) slices at each corresponding $z$-depth are nearly identical.

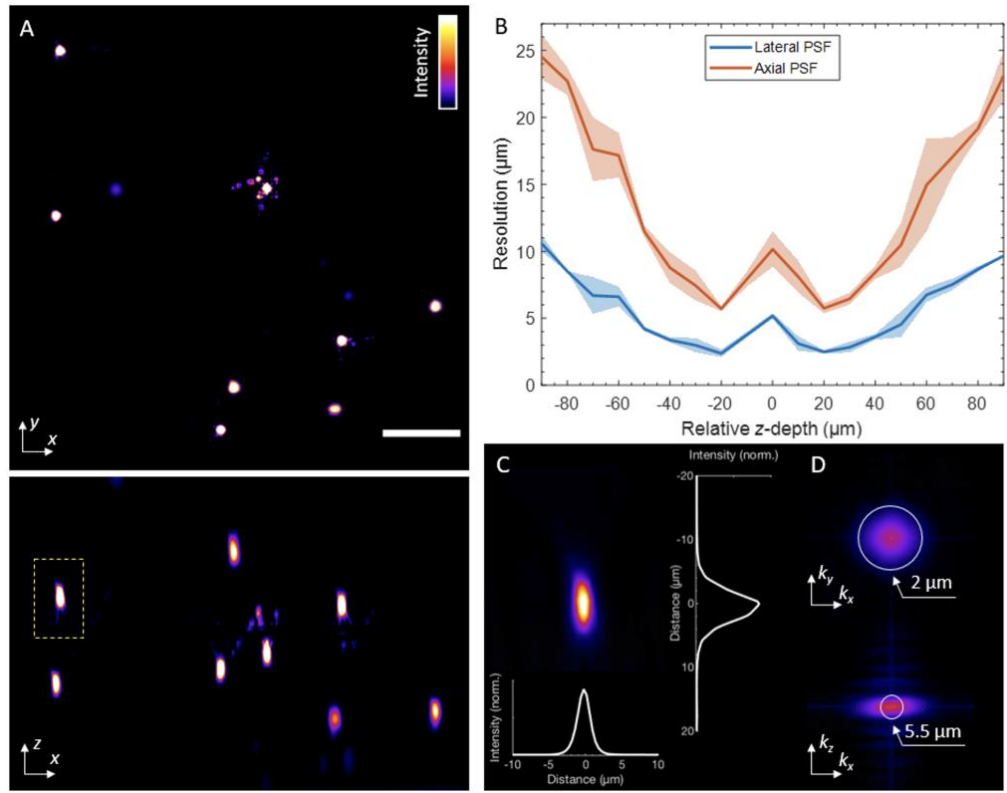

Fig. S3. System resolution. (A) *x-y* (top) and *x-z* (bottom) maximum-intensity projections (MIPs) of a 300- by 300- by 200-μm field of beads in agarose. Scale bar, 50 μm. (B) Lateral (*x-y*) and axial (*x-z*) PSF measurements across the imaging volume, where $z = 0$ is the native focus ($N =$ 53 FWHM bead images at different depths). Means (center lines) and standard deviations (shadings) are shown. (C) Enlarged view of the *x-z* MIP of an exemplary bead from the image volume, denoted by dashed yellow rectangle (in A, bottom). Line profiles of the lateral (bottom) and axial (right) intensities through the center of the bead. (D) Same bead presented as lateral (top) and axial views of the optical transfer functions with resolution bands (white circles).

## 2. *k*-space filtering

We describe here our *k*-space filtering process to alleviate light-field microscopy (LFM) reconstruction artifacts. These grid-like artifacts are due to the degeneracy in spatio-angular sampling at the native focal plane, and have been described theoretically and experimentally [4]. Our method is motivated by two empirical observations. First, the grid-like artifacts are mainly composed of spatial frequencies beyond the theoretical resolution limit of the detection optics (Fig. S5A, left column). Second, the artifacts are most prominent at the native focal plane and the immediate axial range around it (Fig. S5B, left column and Fig. S6C). With these observations in mind, we devised the following filtering procedure that selectively removes the bulk of reconstruction artifacts without compromising the resolution of the 3D volume.

At the native focal plane, the theoretical maximum lateral resolution is determined by the diffraction-limited sampling rate of LA: the lenslet pitch divided by the effective magnification [4], which we experimentally confirmed (theory: 5.7 μm; experiment: 5.2 ± 0.2 μm). This resolution limit sets a cutoff frequency in Fourier space (*k*-space) where we can impose a low-pass filter to remove high-frequency noise, the main source of the image artifacts (Fig. S5A, left column). We apply this low-pass filter to the native focal plane and adjacent planes extending across a 10-μm depth, a small subvolume defined by the experimental axial PSF (see dashed yellow rectangles in Fig. S5B). Image planes outside of this subvolume are not low-pass filtered. Note that in LFM the resolution changes as a function of depth, and maximum resolution is achieved at *z* positions away from the native focal plane [4], as experimentally shown in Fig. S3B. Because only the subvolume that extends across the focal plane (where artifacts are most prominent) is *k*-space filtered, higher resolution present elsewhere in the volume is unscathed. Experimental aberrations, background, scattering, and other sources of noise break the underlying assumptions in the reconstruction [1,4], generally decreasing the highest non-zero spatial frequency achievable (i.e., the *effective* resolution limit)—or artificially increasing it—making our *k*-space filter a conservative approach. Our filtering process is outlined in Fig. S4 and can be combined with any LFM reconstruction algorithm.

To quantitatively assess how well *k*-space filtering mitigates reconstruction artifacts, we compared standard LFM and *k*-space filtered reconstructions of a 300- by 200- by 200-μm field of beads in agarose (Fig. S5). In large part the field of beads are similar, but it's clear that artifacts are visible both in lateral and axial maximum-intensity projection (MIP) views of the conventional reconstruction that are not apparent with *k*-space filtering (Fig. S5A). Even though the periodic artifacts are only concentrated at the native focus (Fig. S5B, left column), they persist and lift the noise floor throughout the lateral MIP view (Fig. S5A, left column). High-frequency artifacts can swamp the signal intensity of weak point sources, making it difficult to differentiate artifacts from real signal; in contrast, the *k*-space filtered signal intensities are weighted as expected—where real point sources are located (Fig. S5D, line 2). In addition, filtering significantly decreases reconstruction artifacts without any loss of spatial resolution throughout the 3D volume, as measured by line cuts through several PSFs (Fig. S5D, line 1).

We further tested *k*-space filtering *in vivo*, where background and noise can critically affect the reconstruction quality [1]. We acquired volumetric data of transgenic zebrafish embryos expressing green fluorescent protein in the cranial vasculature by means of LFM and light-sheet microscopy (also known as selective-plane illumination microscopy; SPIM), which provided an additional ground truth (higher resolution) structural image to compare our filtering method against (same dataset as Fig. 2). When applied to living tissue, we observe a dramatic reduction in grid-like artifacts at the native focal plane compared to conventional LFM reconstruction (i.e., no filter), as shown in Fig. S6C. Comparing volumetric contrast in standard and *k*-space filtered reconstruction, we see a dip near the native focal plane (Fig. S6B). This is to be expected, as the grid-like patterns lead to an artificial increase in contrast. Similar to the experimentally measured PSFs, line intensity profiles along filtered blood vessels show an important decrease in spurious spatial signal without loss of resolution (Fig. S6D), alteration of structural features, or additional artifacts (Fig. S6F).

## Light-field reconstruction

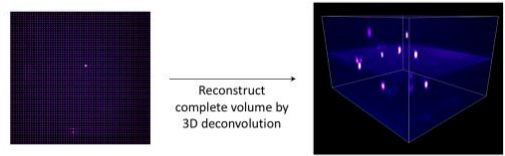

## *k*-space filtering

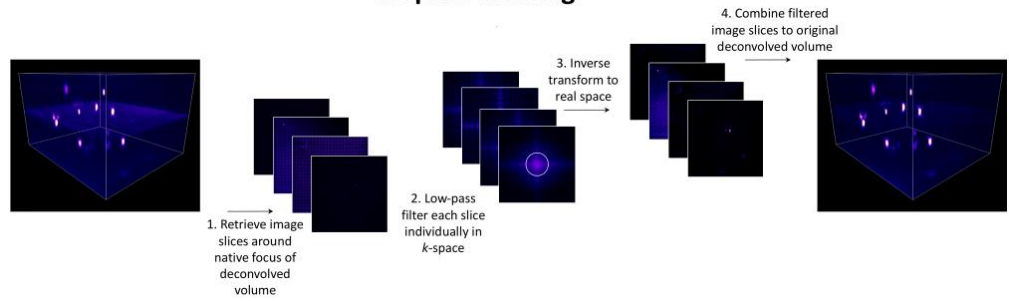

Fig. S4. *k*-space filtering algorithm. LFM (top) reconstructs a complete 3D volume with depth-dependent resolution and artifacts near the native focal plane [4,5]. Due to the non-uniform resolution across the entire volume, a single cutoff frequency cannot be applied without compromising peak resolution at other image planes. *k*-space filtering (bottom) splits the deconvolved volume into smaller subvolumes, and independently processes the subvolume that extends across the native focal plane. Retrieved image slices are low-pass filtered in Fourier (*k*) space, based on the experimental optical transfer function (OTF) bounds at that subvolume. Next, image slices are inverse transformed back into real space, and a median filter is applied to minimize ringing artifacts. The filtered image slices are then combined to assemble the final, denoised volume.

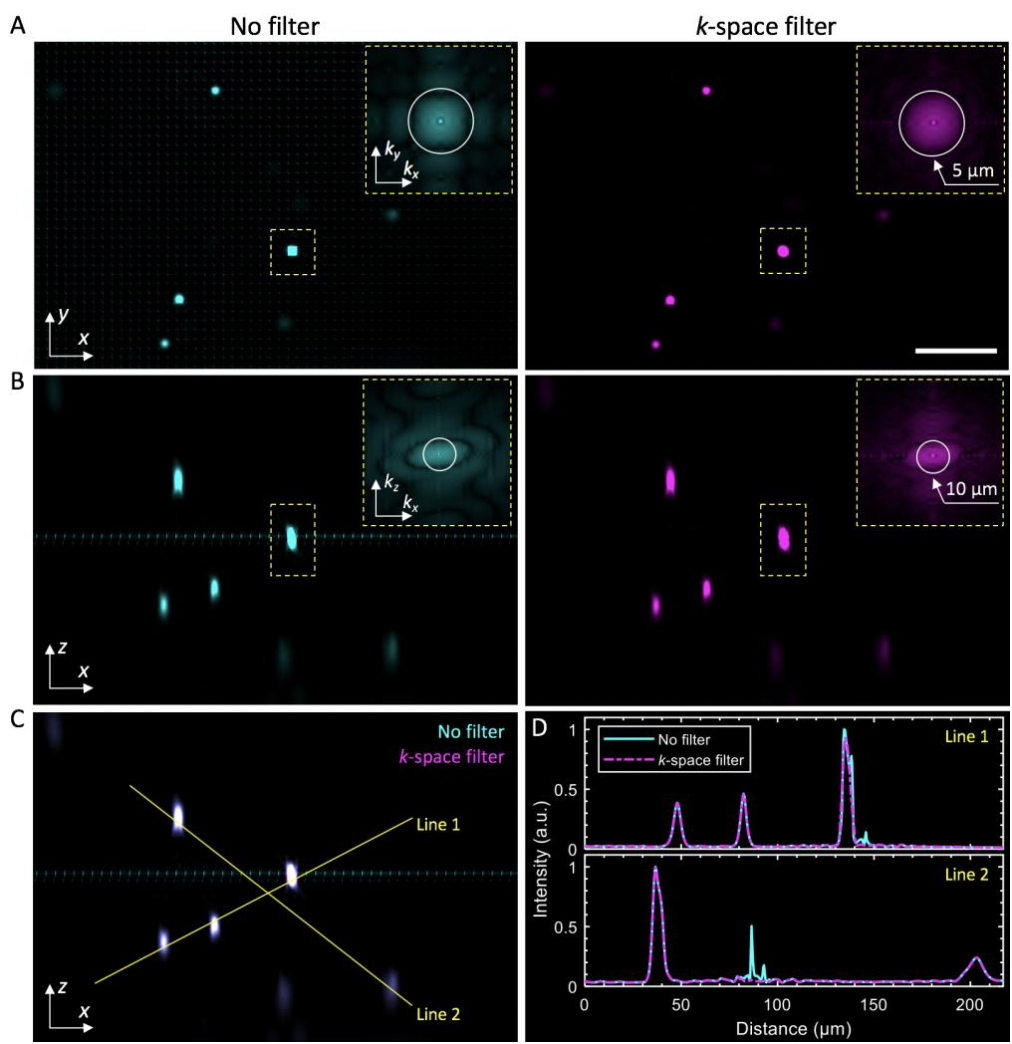

Fig. S5. *k*-space filtering reduces artifacts and improves volume reconstruction with uncompromised resolution. Comparative *x-y* (A) and *x-z* (B) MIPs of a 300- by 200- by 200-μm sparse field of fluorescent beads before (left column) and after filtering in *k*-space (right column). Each inset shows the spatial frequency content of the corresponding axially-centered PSF at the native focus, as indicated by the dashed yellow rectangle in the image. Both real and frequency space representations show the ability of *k*-space filtering to reduce high-frequency artifacts, laterally and axially. OTF images were equally gamma-contrast-adjusted to aid in visualizing weak features. Scale bar, 50 μm. (C) Overlap of *x-z* MIPs show excellent spatial correspondence of PSFs before and after filtering. (D) Comparative line profiles as indicated by the yellow lines in (C). As expected, there is no appreciable loss of resolution by *k*-space filtering (line 1). Away from the native focus, bead-measured signal intensities show full quantitative correspondence, while at the native focal plane, periodic reconstruction artifacts are effectively suppressed (line 2).

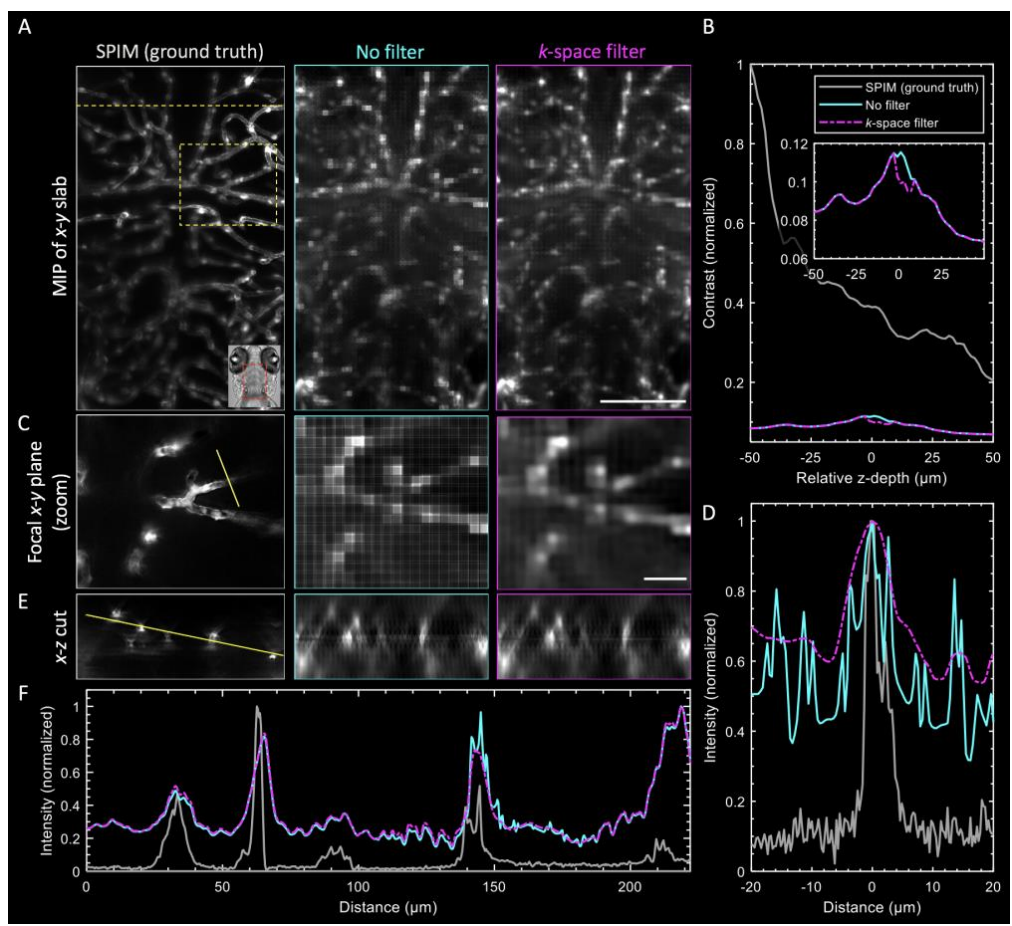

Fig. S6. *k*-space filtering improves volume reconstruction quality *in vivo*. (A) *x-y* MIP of a 100-μm-thick slab (same dataset as Fig. 2), comparing each protocol as shown. Scale bar, 100 μm. (B) Quantification of image contrast versus *z*-depth; each *x-y* slice (from all protocols) was normalized against the deconvolved SPIM (gray) slice at $z = -50$ μm. Inset shows light-field protocols only, with the expected decrease in artificially high contrast by *k*-space filtering. (C) Focal *x-y* plane (zoom) of yellow boxed region in (A), showing a significant decrease in common reconstruction artifacts by *k*-space filtering (third column). Scale bar, 25 μm. (D) Comparative intensity profiles for each protocol, indicated by line cut in (C). (E) *x-z* slice, at the location indicated by the dashed yellow line in the MIP slab in (A). (F) Comparative intensity profiles for each protocol, indicated by the 225-μm yellow line in (E).

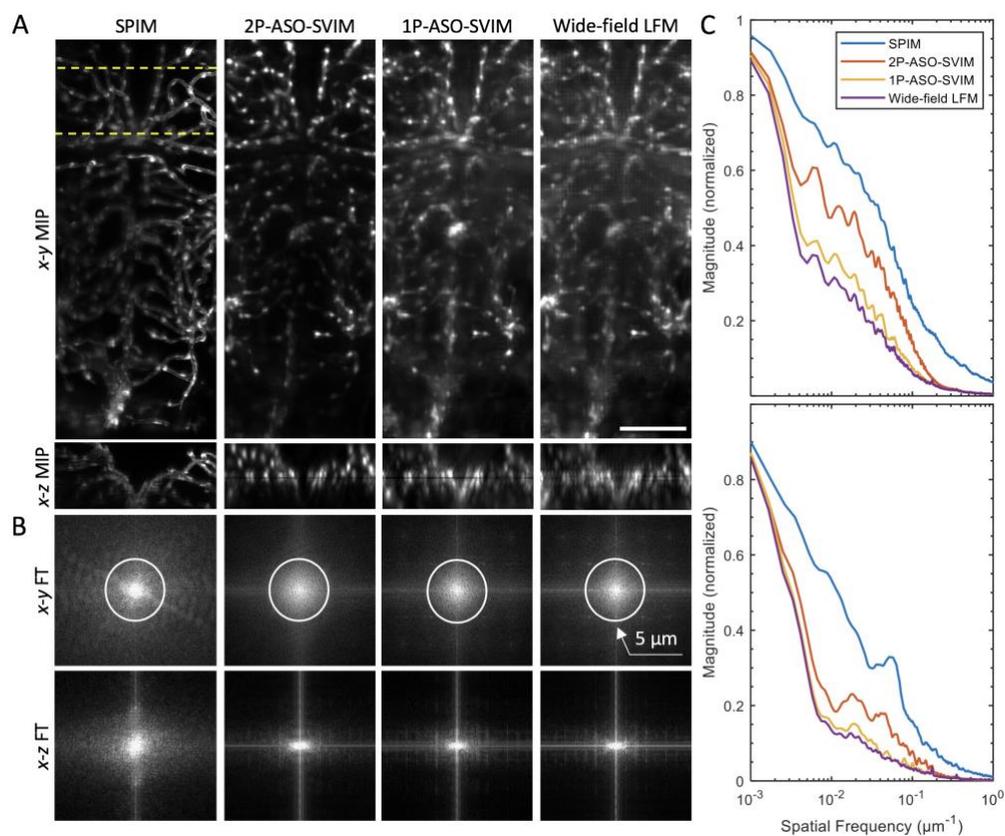

Fig. S7. ASO-SVIM enhances effective resolution across large tissue volumes. (A) *x-y* (top row) and *x-z* (bottom row) MIPs of a 100-μm thick slab (same dataset as Fig. 2), highlighting maximum attenuation for each modality shown. Scale bar, 100 μm. (B) Fourier transforms (FTs) of the MIPs in (A). Resolution bands (white circles) indicate increased spatial frequency content with ASO-SVIM compared to wide-field illumination, due to decreased out-of-volume background by selective excitation. (C) Average amplitudes along the $k_y$ (top) and $k_z$ direction (bottom) of FTs in (B), respectively. Frequency spectra demonstrate the slower spatial frequency roll-off for ASO-SVIM, both laterally and axially, and hence improved effective 3D resolution over the conventional wide-field technique. See also Fig. 2 and Fig. S8.

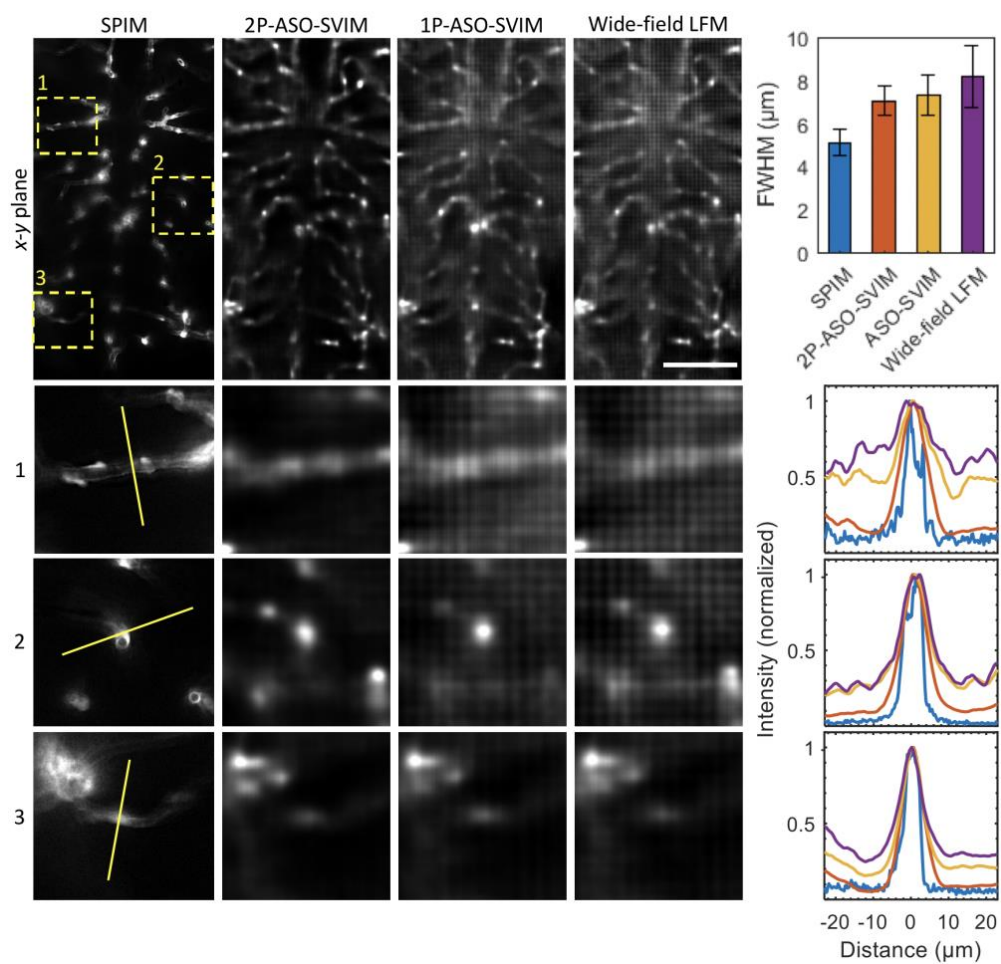

Fig. S8. Comparison of line cuts through vessel structures *in vivo*. Top row: *x-y* slice from a 100-μm thick slab (same dataset as Fig. 2), centered at approximately 86 μm into the specimen (z = -14 μm), comparing the performance of the indicated modalities. Remaining rows: Zoomed-in regions of structures in the yellow boxes in the *x-y* plane (top row), along with corresponding line intensity profiles (as shown by the 50-μm yellow line in the images) plotted on the right. Given the intrinsically higher spatial resolution of SPIM, full quantitative correspondence of the light-field-based images is not expected. All 3-line profiles were used to quantify the average FWHM and standard deviation for each modality (right column, top). Of the light-field-based methods, 2P-ASO-SVIM achieves the highest biological resolution (owing to nonlinear excitation as well as reduced background and scattering), approaching the performance of SPIM, followed by ASO-SVIM in 1P mode, and last, wide-field LFM. Scale bar, 100 μm. See also Fig. 2 and S7.

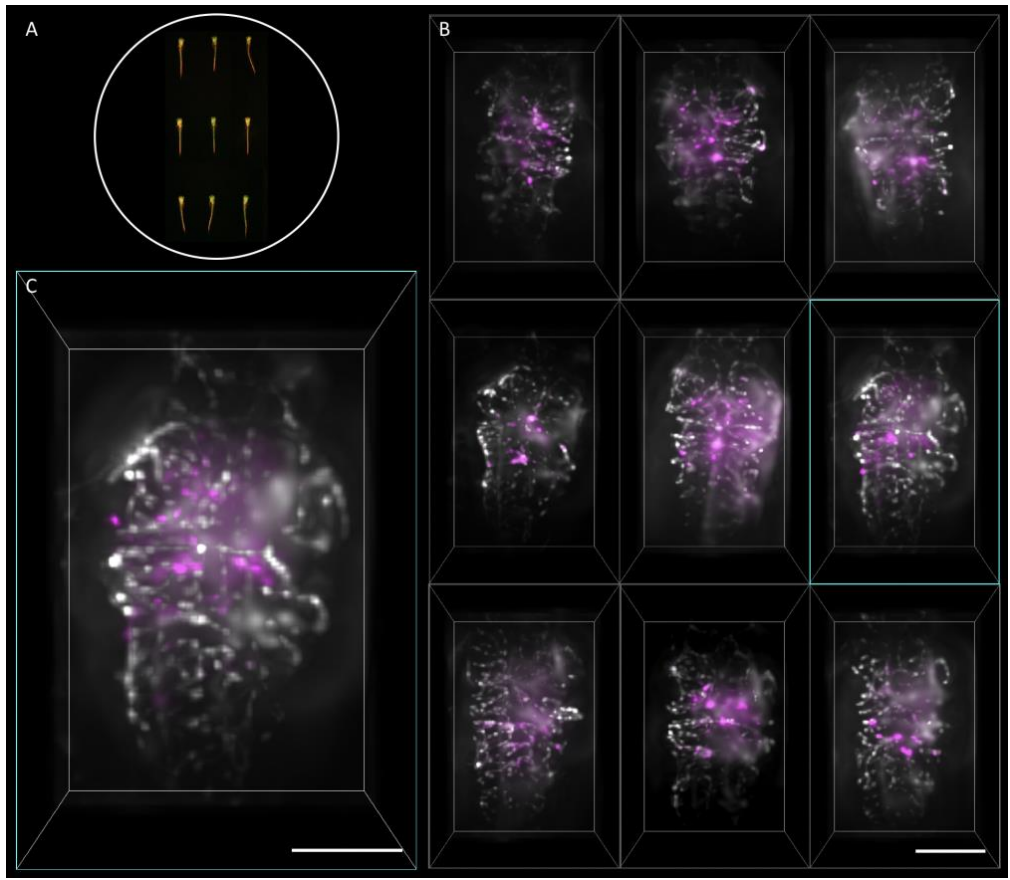

Fig. S9. ASO-SVIM enables high-throughput imaging of whole-brain blood flow in zebrafish larvae. (A) Nine 5-dpf zebrafish, with fluorescent labels in both the blood cells [Tg(gata1:dsRed)] and endocardium [Tg(kdrl:eGFP)], were mounted in a standard multi-well plate. Two-color imaging was performed over a synchronous 670- by 470- by 200-μm volume at ~50 Hz, per color, for each fish. (B) MIPs of 50-μm-thick slabs axially-centered within the 9 dually-labeled zebrafish brains. Captured light-fields were reconstructed using ray optics [3] for increased computational speed. Blood cells and endocardium are represented in magenta and grayscale, respectively. Scale bar, 200 μm. (C) Magnified MIP of the specimen highlighted by the cyan box in (B). Scale bar, 100 μm. See Visualizations 2-4.

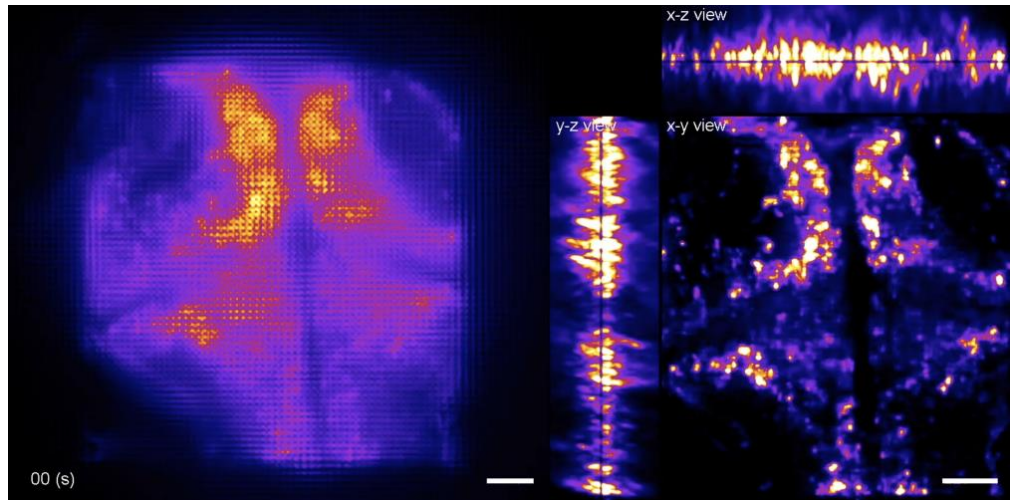

Visualization 1. Fluorescence light-field (left) and 3D reconstructed maximum-intensity projections along the indicated directions (right) of a time-lapse recording of brain-wide neural activity in a 5-dpf transgenic zebrafish. 2P-ASO-SVIM imaging was performed at a volumetric rate of 1 Hz. Same dataset as presented in Fig. 3. Scale bar (left), 150 µm, (right) 50 µm.

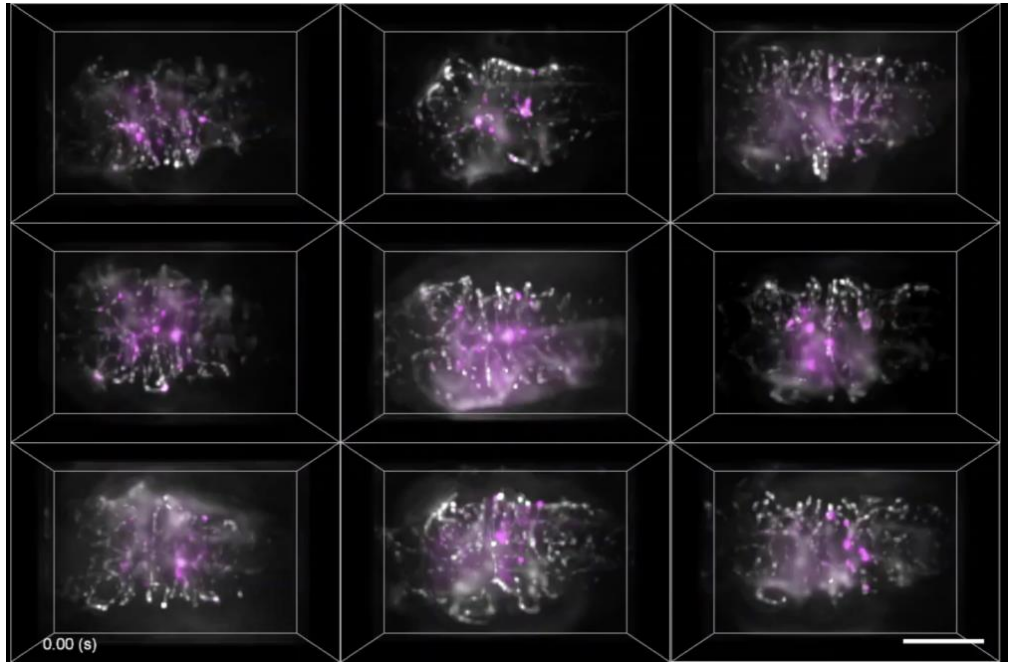

Visualization 2. Maximum-intensity projections of nine 5-dpf zebrafish, with fluorescent labels in both the blood cells [Tg(gata1:dsRed)] and endocardium [Tg(kdrl:eGFP)], represented in magenta and grayscale, respectively. Samples were recorded serially, with each sample imaged with 1P-ASO-SVIM over a synchronous 670- by 470- by 200-μm volume at ~50 Hz. Same dataset as presented in Fig. S9. Scale bar, 200 μm.

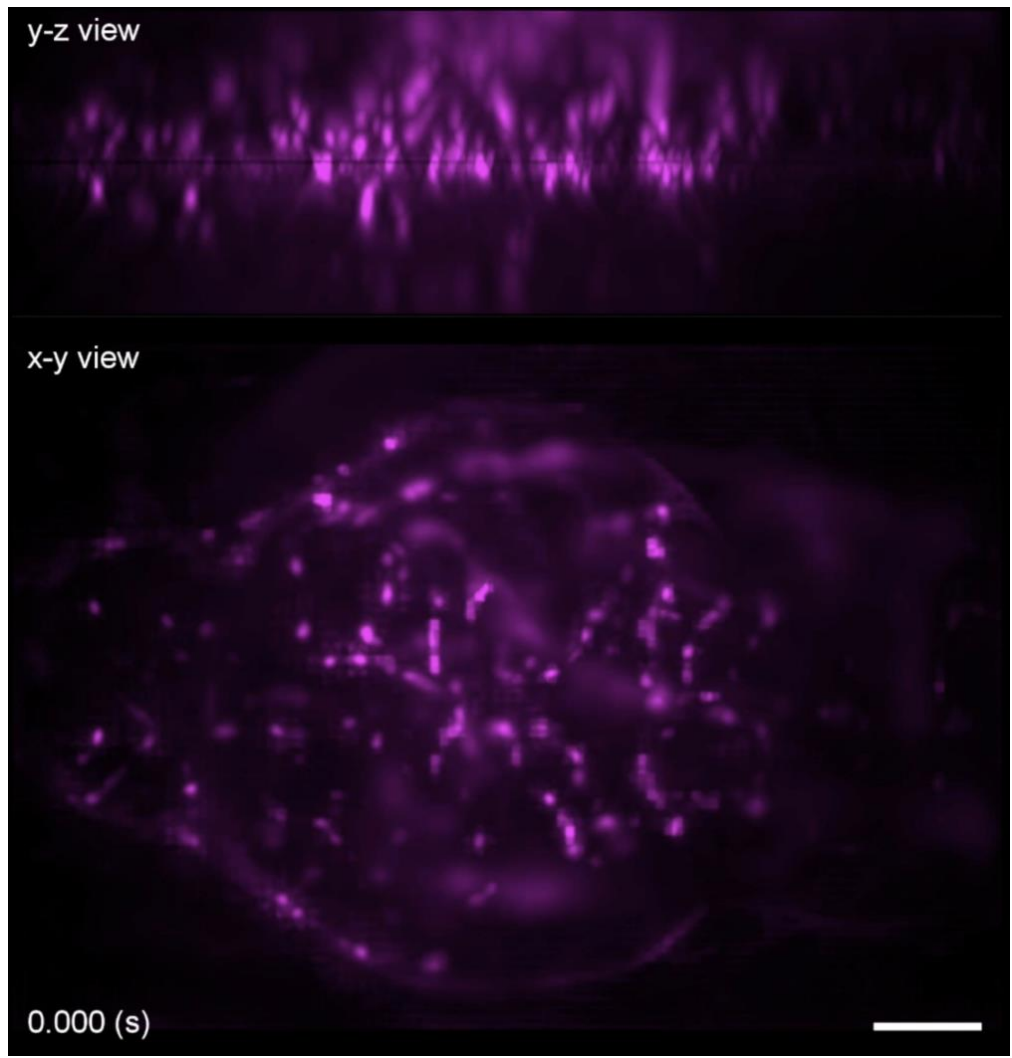

Visualization 3. 1P-ASO-SVIM imaging of blood cells [Tg(gata1:dsRed)] flowing across the entire brain of a 5-dpf zebrafish. Cellular resolution imaging was performed over a 670- by 470- by 200-μm volume at ~50 Hz. Fluorescent light-fields were wave optics reconstructed. Animal is oriented anterior (left) to posterior (right). Scale bar, 50 μm.

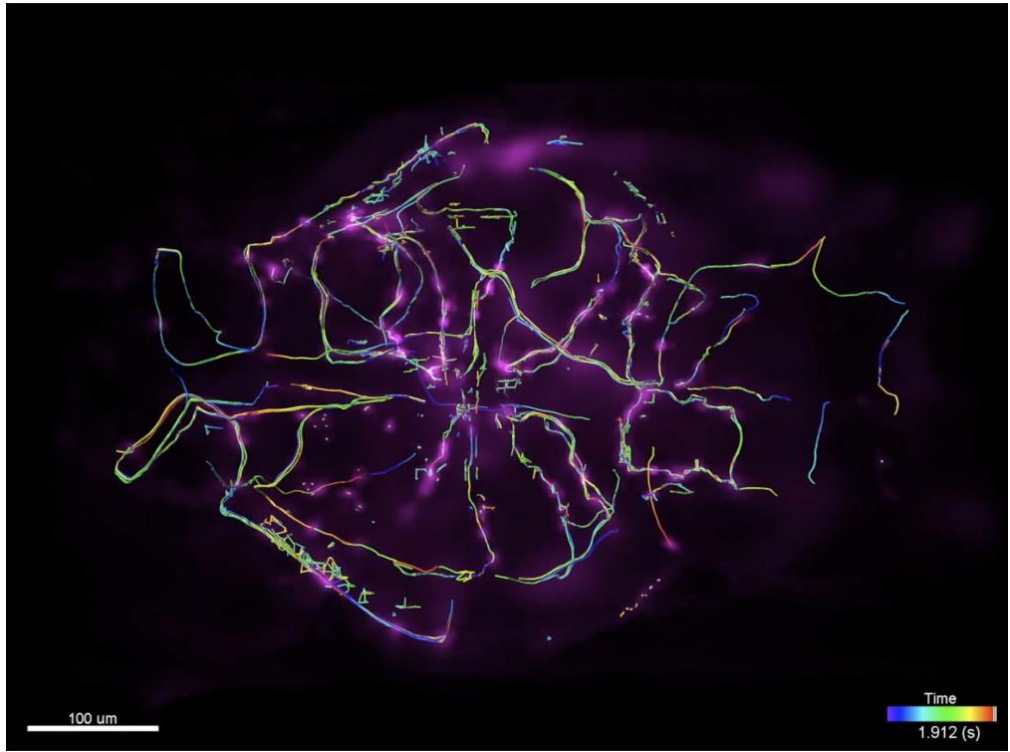

Visualization 4. Volumetric view of whole-brain blood flow, with red blood cell tracks color-coded in time. Same dataset as presented in Visualization 3.

**References**

1. T.V. Truong, D.B. Holland, S. Madaan, A. Andreev, K. Keomanee-Dizon, J.V. Troll, D.E.S. Koo, M.J McFall-Ngai, and S.E. Fraser, "High-contrast, synchronous volumetric imaging with selective volume illumination microscopy," Communications Biology **3**, 1-8 (2020).
2. K. Keomanee-Dizon, S.E. Fraser, and T.V. Truong, "A versatile, multi-laser twin-microscope system for light-sheet imaging," Review of Scientific Instruments **91**, 053703 (2020).
3. M. Levoy, R. Ng, A. Adams, M. Footer, and M. Horowitz, "Light field microscopy," in *Proceedings of ACM SIGGRAPH* (ACM, 2006) p. 924-934.
4. M. Broxton, L. Grosenick, S. Yang, N. Cohen, A. Andalman, K. Deisseroth, and M. Levoy, "Wave optics theory and 3-D deconvolution for the light field microscope," Optics Express **21**, 25418-25439 (2013).
5. R. Prevedel, Y. G. Yoon, M. Hoffmann, N. Pak, G. Wetzstein, S. Kato, T. Schrödel, R. Raskar, M. Zimmer, E. S. Boyden, and A. Vaziri, "Simultaneous whole-animal 3D imaging of neuronal activity using light-field microscopy," Nature Methods **11**, 727-730 (2014).
6. T.A. Pologruto, B.L. Sabatini, and K. Svoboda, "ScanImage: flexible software for operating laser scanning microscopes," Biomedical engineering online **2**, 13 (2003).
7. A. D. Edelstein, M. A. Tsuchida, N. Amodaj, H. Pinkard, R. D. Vale, and N. Stuurman, "Advanced methods of microscope control using μManager software," Journal of Biological Methods **1**, 10 (2015).